\documentclass[nofootinbib,notitlepage,superscriptaddress,10pt,aps,pra,twocolumn]{revtex4-1}

\usepackage[utf8x]{inputenc}
\usepackage{amsmath}
\usepackage{amssymb}
\usepackage{graphicx}
\usepackage[normalem]{ulem}
\usepackage{epsf,epsfig}
\usepackage{psfrag}
\usepackage{color}
\usepackage{multirow}
\usepackage{diagbox}

\makeatletter
\newcommand\stroke[1]{\mathpalette\stroke@aux{#1}}
\def\stroke@aux#1#2{%
  \ooalign{%
    \hfil$#1^{\;\, \_\hspace{-0.05cm}\_}$\hfil\cr
    \hfil$#1#2$\hfil\cr
  }%
}
\makeatother

\begin{document}

\title{In-vacuo-dispersion-like spectral lags
in gamma-ray bursts}

\author{Giovanni Amelino-Camelia}
\affiliation{Dipartimento di Fisica, Universit\`a di Roma ``La Sapienza", P.le A. Moro 2, 00185 Roma, Italy}
\affiliation{INFN, Sez.~Roma1, P.le A. Moro 2, 00185 Roma, Italy}
\author{Giacomo D'Amico}
\affiliation{Dipartimento di Fisica, Universit\`a di Roma ``La Sapienza", P.le A. Moro 2, 00185 Roma, Italy}
\affiliation{INFN, Sez.~Roma1, P.le A. Moro 2, 00185 Roma, Italy}
\author{Fabrizio Fiore}
\affiliation{Osservatorio Astronomico di Roma,
Via Frascati 33, Monte Porzio Catone, I00078, Italy}
\author{Simonetta Puccetti}
\affiliation{Agenzia Spaziale Italiana,
Unit\`a di Ricerca Scientifica,
Via del Politecnico, 00133 Roma, Italy}
\author{Michele Ronco}
\affiliation{Dipartimento di Fisica, Universit\`a di Roma ``La Sapienza", P.le A. Moro 2, 00185 Roma, Italy}
\affiliation{INFN, Sez.~Roma1, P.le A. Moro 2, 00185 Roma, Italy}

\begin{abstract}
Some recent studies exposed rather strong statistical evidence of in-vacuo-dispersion-like spectral lags
for gamma-ray bursts (GRBs), a linear correlation between time of observation and energy of GRB particles.
Those results focused on testing in-vacuo dispersion for the most energetic GRB particles, and in particular
 only included  photons with energy
at emission greater than 40 GeV.
We here extend the window of the statistical analysis down to 5 GeV and find results
that are consistent with what had been previously noticed at higher energies.
\end{abstract}
\maketitle

Over the last 15 years there has been considerable interest
(see {\it e.g.}
Refs.\cite{gacLRR,jacobpiran,gacsmolin,grbgac,gampul,urrutia,gacmaj,myePRL,gacGuettaPiran,steckerliberati} and references therein)
in quantum-gravity-induced
in-vacuo dispersion, the possibility that spacetime itself might behave essentially like a dispersive medium
for particle propagation: there might be
an energy dependence of the travel times
of ultrarelativistic particles
  from a given source to a given detector.

It is well established \cite{gacLRR,jacobpiran,gacsmolin,grbgac} that the analysis of GRBs could
allow us to test this in-vacuo-dispersion hypothesis.
Some of us were involved in the first studies
using IceCube data for searching for GRB-neutrino in-vacuo-dispersion
candidates~\cite{gacGuettaPiran,Ryan,RyanLensing,gacnatureastronomy}.
Analogous investigations were performed
in a series of studies \cite{MaZhang,MaXuPRIMO,MaXuSECONDO} (also see \cite{gacFioreGuettaPuccetti})
focusing on the highest-energy GRB photons observed by the Fermi telescope.
As summarized in Fig.1 these studies provided rather strong statistical evidence of
in-vacuo-dispersion-like spectral lags. For each point in Fig.1 we denote
by  $\Delta t$ the difference between the time of observation of the relevant particle
and the time of observation of the first low-energy peak in  the GRB,
while
$E^*$ is the redshift-rescaled energy of the relevant particle:
\begin{equation}
E^* \equiv E \frac{D(z)}{D(1)}
\label{tstar}
\end{equation}
where $z$ is the redshift of the relevant GRB and
\begin{equation}
D(z) = \int_0^z d\zeta \frac{(1+\zeta)}{H_0\sqrt{\Omega_\Lambda + (1+\zeta)^3 \Omega_m}}  \, .
\label{dz}
\end{equation}
$\Omega_\Lambda$, $H_0$ and $\Omega_m$ denote, as usual,
respectively the cosmological constant, the Hubble parameter and the matter fraction,
for which we take the values given in Ref.\cite{PlanckCosmPar}.

\begin{figure}
\includegraphics[scale=0.75]{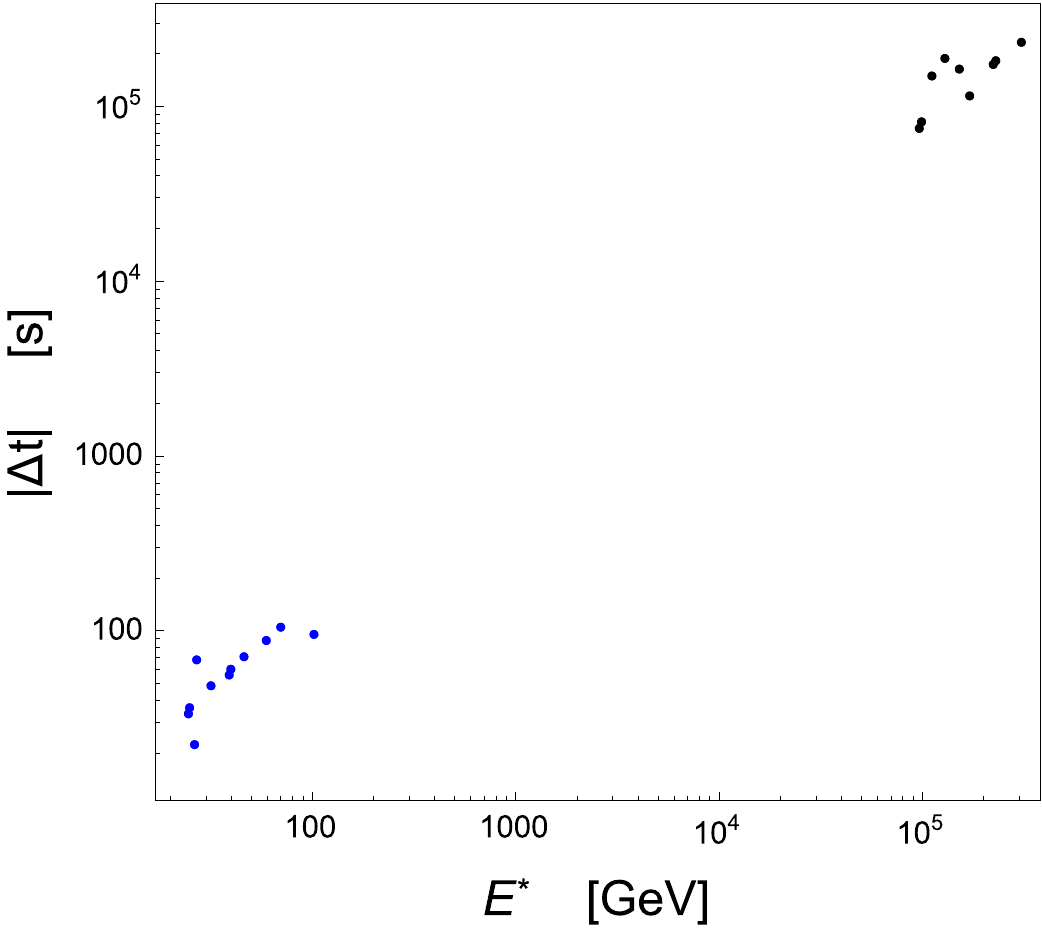}
\caption{Values of $|\Delta t|$ versus $E^*$ for the
IceCube  GRB-neutrino
candidates discussed in Refs.~\cite{Ryan,gacnatureastronomy} (black points)
and for the GRB photons discussed in
Refs.\cite{MaXuSECONDO,gacnatureastronomy} (blue points).
The photon points in  figure also factor in
the result of a one-parameter fit estimating the average magnitude of intrinsic time lags (details in Refs.\cite{MaXuSECONDO,gacnatureastronomy}).}
\end{figure}

The most studied
\cite{gacLRR,jacobpiran,gacsmolin,grbgac,gampul,urrutia,gacmaj,myePRL,gacGuettaPiran,steckerliberati}
modelization of quantum-gravity-induced in-vacuo dispersion is
\begin{equation}
\Delta t = \eta_X \frac{E}{M_{P}} D(z) \pm \delta_X \frac{E}{M_{P}} D(z) \, ,
\label{main}
\end{equation}
which in terms of $E^*$ takes the form
\begin{equation}
\Delta t = \eta_X D(1) \frac{E^*}{M_{P}}  \pm \delta_X D(1) \frac{E^*}{M_{P}}  \, .
\label{maintwo}
\end{equation}
$M_{P}$ denotes the Planck scale ($\simeq 1.2\,\cdotp 10^{28}eV$) and
the values of the  parameters $\eta_X$ and $\delta_X$ in (\ref{main})
are to be determined experimentally. In (\ref{main}) the notation ``$\pm \delta_X$"
reflects the fact that $\delta_X$ parametrizes the size of quantum-uncertainty (fuzziness) effects. Instead the parameter $\eta_X$
characterizes systematic effects: for example in our conventions for positive $\eta_X$ and $\delta_X =0$ a high-energy particle
is detected systematically after a low-energy particle (if the two particles are emitted simultaneously).
The label $X$ for $\delta_X$ and $\eta_X$
intends to allow for a possible dependence \cite{gacLRR,steckerliberati}
of these parameters on the type of particles (so that for example for neutrinos and photons one would have
$\eta_\nu$, $\delta_\nu$, $\eta_\gamma$, $\delta_\gamma$) and in principle also on spin/helicity
(so that for example for neutrinos one would have
$\eta_{\nu +}$, $\delta_{\nu +}$, $\eta_{\nu -}$, $\delta_{\nu -}$).

The black points in Fig.1
are  ``GRB-neutrino candidates" in the sense
of Ref.~\cite{Ryan}, while the blue points correspond to GRB photons with energy at emission
greater than 40 GeV. The linear correlation between $\Delta t$ and $E^*$ visible in Fig.1 is just of
the type expected for quantum-gravity-induced in-vacuo dispersion.  It might of course be accidental,
but it has been estimated~\cite{Ryan} that for the relevant GRB-neutrino candidates such a high level of correlation would occur accidentally only in less than $1\%$ of
cases, while GRB photons could produce
such high correlation (in absence of in-vacuo dispersion)
only in less than $0.1\%$ of cases~\cite{gacnatureastronomy}. The ``statistical evidence"
summarized in Fig.1
is evidently intriguing enough to motivate us to explore whether
or not the in-vacuo-dispersion-like spectral lags persist at lower energies.

One challenge for this is that evidently
we cannot simply apply to lower-energy photons the reasoning which led to Fig.1:
as stressed above the $\Delta t$ in Fig.1
is the difference between the time of observation of the relevant particle
and the time of observation of the first low-energy peak in  the GRB,
so it is a $\Delta t$ which makes sense for in-vacuo-dispersion studies only
for photons which one might think were emitted in (near) coincidence with the
first peak of the GRB. This assumption is (challengeable \cite{ghisellini} but) plausible~\cite{MaXuSECONDO}
for the few highest-energy GRB photons relevant for Fig.1, with energy at emission greater than 40 GeV,
but of course it cannot apply to all photons in a GRB. Conceptually the main aspect of novelty
of our analysis concerns a strategy for handling this challenge.

We consider the same GRBs relevant for the analysis
summarized in Fig.1, but now including all photons from those GRBs with energy at the source
greater than 5 GeV, thereby lowering the cutoff by nearly an order of magnitude. Only 11 photons
took part in the previous analyses whose findings were summarized in our Fig.1, whereas the analysis we are here reporting involves a total of 148 photons.
For the reasons discussed above, we do not consider
the $\Delta t$
 (with reference to the first peak of the GRB), but rather we consider
a $\Delta t_{pair}$, which gives for each pair of photons
in our sample their difference of
time of observation.
Essentially each pair of photons (from the same GRB) in our sample
is taken to give us an estimated value of $\eta_\gamma$,
by simply computing
\begin{equation}
\eta_\gamma^{[pair]} \equiv \frac{M_{P} \Delta t_{pair}}{D(1) E^*_{pair}} \, ,
\label{etapair}
\end{equation}
where $E^*_{pair}$ is the difference in values of $E^*$ for the two photons in the pair.
Of course the $\Delta t_{pair}$ for many pairs of photons in our sample could  not possibly have anything
to do with in-vacuo dispersion: if the two photons were produced from different phases
of the GRB (different peaks) their $\Delta t_{pair}$ will be dominated by the intrinsic
time-of-emission difference. Those values of $\eta_\gamma^{[pair]}$ will be spurious, they will be ``noise" for
our analysis. However we also of course expect that some pairs of photons in  our sample
were emitted nearly simultaneously,
and for those pairs the $\Delta t_{pair}$ could truly estimate $\eta_\gamma$.
Since estimating $\eta_\gamma$ from the photons
in Fig.1 one gets $\eta_\gamma = 30 \pm 6$, the preliminary
evidence here summarized in Fig.1 would find additional support if this sort of analysis
showed that values of $\eta_\gamma^{[pair]}$ of  about 30 are surprisingly frequent, more frequent
than expected without
 a relationship between arrival times and energy of the type  produced by in-vacuo dispersion.

\begin{figure}
\includegraphics[scale=0.6]{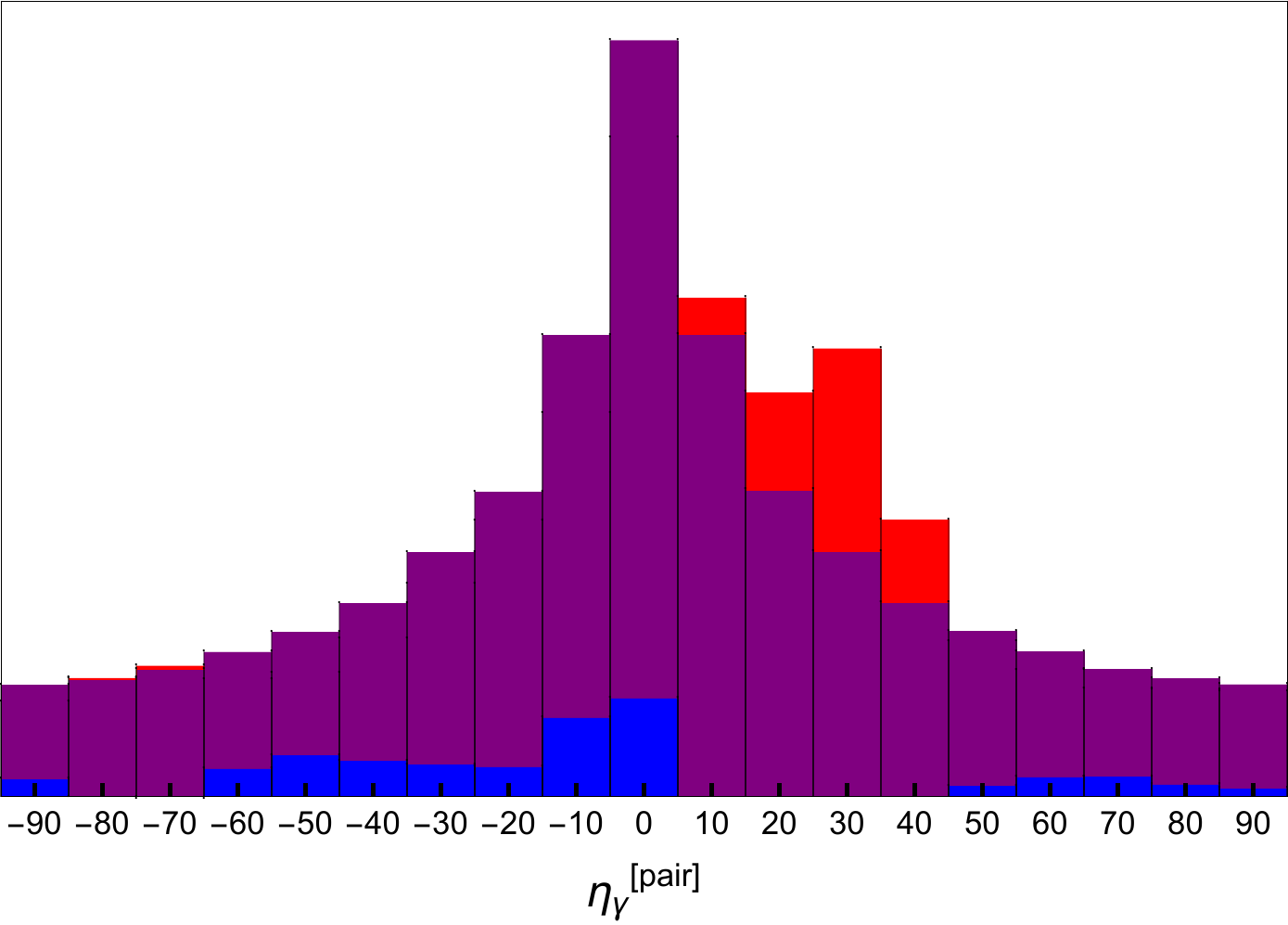}
\caption{Normalized distribution of $\eta_\gamma^{[pair]}$ for all
pairs of photons (from the same GRB) within our data set.
 For bins where the observed population is higher than expected we color the bar in purple up to the level expected, showing then the excess in  red.
For bins where the observed population is lower than expected
the bar height gives the expected population, while the blue portion of the bar quantifies the amount by which the observed population is lower than expected.}
\end{figure}

This is just what we find, as shown perhaps most vividly by the content of Fig.2.
The main point to be noticed
in Fig.2 is that we find in our sample a frequency of occurrence of values of $\eta_\gamma^{[pair]}$
between 25 and 35 which is tangibly higher than one  would have expected in absence of a correlation
between $\Delta t_{pair}$ and $E^*_{pair}$.
Following a standard strategy of analysis (see, {\it e.g.},  Refs.\cite{gacFioreGuettaPuccetti,prdvasil})
we estimate how frequently $25 \leq \eta_\gamma^{[pair]} \leq 35$
should occur in absence of correlation
between $\Delta t_{pair}$ and $E^*_{pair}$ by producing $10^5$ sets of simulated data,
each obtained by reshuffling randomly the times of observation of the photons in our sample.
More details on this and other aspects of our analysis are given in Appendix A.
Also in Appendix A we show that our findings are strikingly robust with respect to restricting
the analysis to only part of our data set:
values of $\eta_\gamma^{[pair]}$
between 25 and 35 occur at a rate higher than expected for all meaningful portions of our data set.
Most notably, values of $\eta_\gamma^{[pair]}$
between 25 and 35 occur at a rate higher than expected even if we exclude from the analysis
the photons whose energy at emission is greater than 40 GeV (the photons that were taken into account
in the analyses leading to the content of our Fig.1).

It is also noteworthy that we find (see Appendix A) that
an excess of results for $\eta_\gamma^{[pair]}$
between 25 and 35 as big as shown by our data
 should occur accidentally (in absence of in-vacuo
dispersion) in less than 0.5$\%$ of cases.

Also intriguing is the content of our Fig.3,
 which offers an intuitive characterization of the consistency that emerged
 from our analysis between what had been found in previous studies of GRB photons with energy at
 emission greater than 40 GeV, and what we now find for GRB photons with energy
 between 5 and 40 GeV.

\begin{figure}
\includegraphics[scale=0.75]{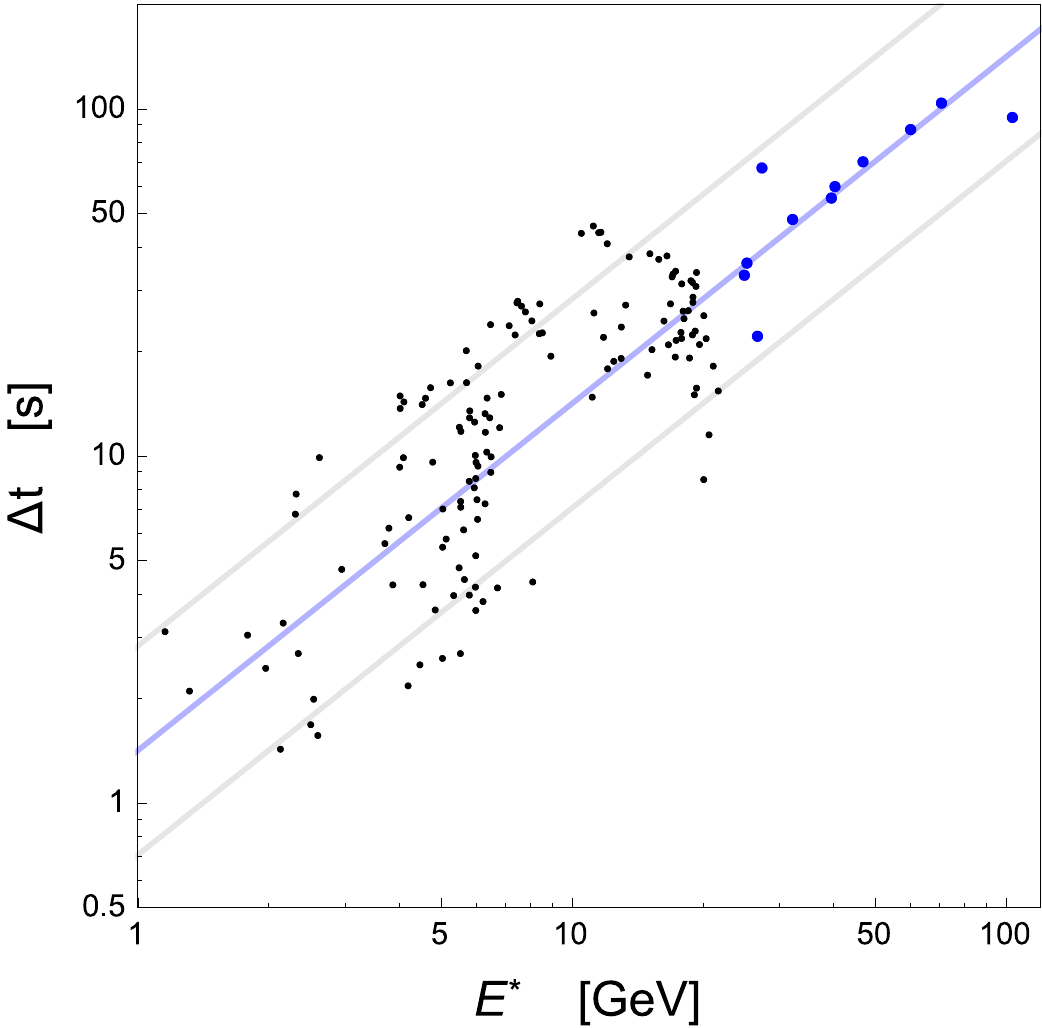}
\caption{As in Fig.1 blue points here are for the GRB photons discussed in
Refs.\cite{MaXuSECONDO,gacnatureastronomy} (with energy at emission greater than 40 GeV).
Here black points give the $E^*_{pair}$  and the $\Delta t_{pair}$  for our pairs of GRB photons,
including only cases in which both photons have energy at emission lower than 40 GeV and the associated
value of $\eta_\gamma^{[pair]}$ is rather sharp (relative error of less than 30$\%$) and between 10 and 100.
The gray lines characterize the range of values of $\eta_\gamma$ favored
by the blue points, which is also the region where black points are denser. 
The violet line is for $\eta_\gamma = 34$ and intends to help the reader notice the similarity of statistical properties between the distribution of black and blue points, that goes perhaps even beyond the quantitative aspects exposed in our histograms.}
\end{figure}

As a further test of robustness of our findings we performed a variant of our analysis
focused on triplets of photons rather than pairs.
For any 3 photons (from the same GRB, of course)
in our data sample we obtain an estimated value of $\eta_\gamma$
by a best-fit technique described in  detail in  Appendix A.
Evidently also for these triplets we expect a combination
of spurious results for  $\eta_\gamma$ (photons in the triplets being emitted
at the times of two or three different peaks of the GRB)
and of meaningful results  for  $\eta_\gamma$ (cases of three photons
emitted nearly simultaneously). As one can easily infer from
Fig.4 our statistics for such triplets is rather low
and as a result our estimate of the expected distribution is  not fully robust, but still
the excess of results for $\eta_\gamma$
between 25 and 35 is so large that we can confidently assume it is meaningful.

\begin{figure}
\includegraphics[scale=0.6]{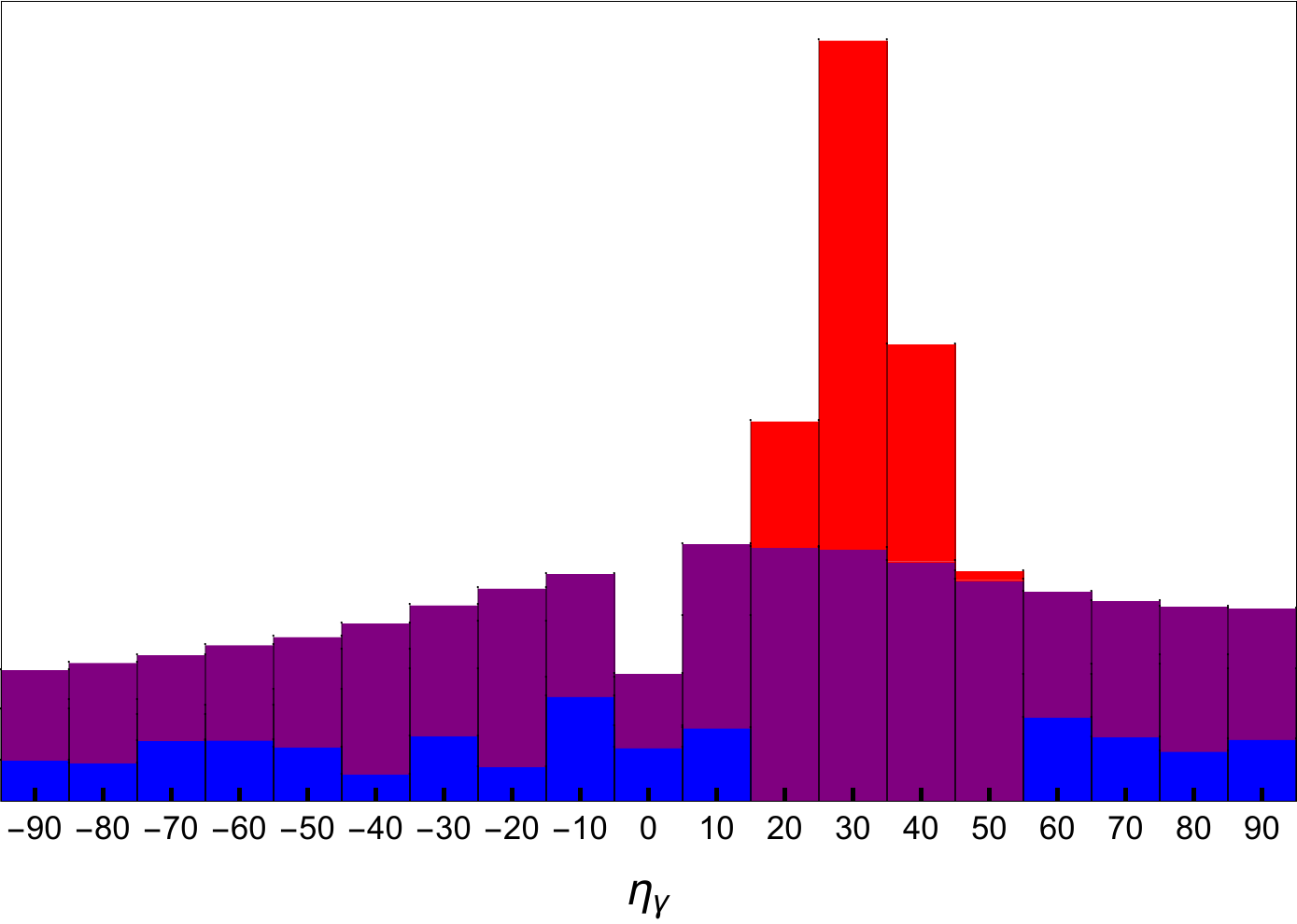}
\caption{Normalized distribution of $\eta_\gamma$ for triplets of photons in our sample.
Same color coding for bars as in Fig.2.}
\end{figure}

In summary  we found rather striking indications in favor
of values of $\eta_\gamma$
of about 30 in GRB data for all photons with energy at emission
greater than 5 GeV. We used data that were already available at the time
of the studies that led to Fig.1 (which in particular focused on photons with energy at
emission greater than 40 GeV) but nobody had looked before
at those data for photons with energy at emission
 between 5 and 40 GeV, from the
 perspective of Fig.1. We therefore feel that it might be legitimate to characterize
what we here reported as a successful prediction originating from the analyses on which
Fig.1 was based. Combining the statistical significance here exposed with the already
noteworthy statistical significance of the independent
analyses~\cite{Ryan,gacnatureastronomy,MaXuSECONDO} whose findings were
here summarized in Fig.1, we are starting to lean toward expecting that  not  all of this
is accidental,  in the sense that on future similar-size GRB data samples one should
find again at least some partial manifestation of
the same feature. We are of course much further from establishing
whether this feature truly is connected with quantum-gravity-induced in-vacuo dispersion,
rather than being some intrinsic property of GRB signals.
Within our analysis the imprint of in-vacuo dispersion is coded in the $D(z)$
for the distance dependence and, while that does give a good match to the data,
one should keep in mind that only a few redshifts (a few GRBs) were relevant
for our analysis. If we are actually seeing some form of in-vacuo dispersion it would
most likely
be of statistical (``fuzzy") nature since other studies have provided evidence strongly disfavoring
the possibility that this type of in-vacuo-dispersion effects
would affect systematically all photons \cite{fermiNATURE}.


\appendix

\section{}
In this appendix we provide further details on the results discussed in the main text
and also discuss some additional corollary results.

Our analysis focuses on the same GRBs whose photons took part in the analyses
which led to the picture here summarized in Fig.1. These are the GRBs that provide
us the full range of energies relevant for our analysis, including some photons
with energy at emission greater than 40 GeV:  {\small GRB080916C, GRB090510, GRB090902B, GRB090926A,
GRB100414A, GRB130427A, GRB160509A}. The relevant data were downloaded from the Fermi-LAT archive 
 and they were calibrated and cleaned using the LAT ScienceTools-v10r0p5 package, which is available from the Fermi Science Support Center.

For reasons that shall soon be clear it was valuable for us to divide our data sample
in different subgroups, characterized by different ranges of values for the energy at
emission, which we denote by $E_0$. We label as ``high" the photons in our sample with  $E_0 > 40 GeV$,
with "medium" those with $15 GeV \leq E_0 \leq 40 GeV$, and with ``low" those with
$5 GeV \leq E_0 \leq 15 GeV$. Our ``high" photons were already taken into account in the previous studies which
led to Fig.1, so it is particularly valuable to keep them distinct from the other photons
in our sample (the ones we label as ``medium" and ``low").

Let us start with the content of Fig.2, which
takes into account all pairs of photons (of course from the same GRB)
 within our data set.
Each such pair typically contributes to more than one  of our bins, considering
that the energies of the photons are not known very precisely.
The contribution of a given pair to each bin is computed generating a gaussian distribution with mean value $\eta_\gamma$ (calculated with Eq. \eqref{etapair}) and standard deviation $\sigma_\gamma$ obtained by error propagation of the energy uncertainty, which we assume to be of 10$\%$. Then, we compute the area of this distribution, which we limit in the interval $[\eta_\gamma -\sigma_\eta , \eta_\gamma + \sigma_\eta]$, falling within each bin, in order to evaluate the value to assign to a given bin. Thus, each pair in general contributes to more than one bin and does that with a gaussian weight.
The expected frequency of occurrence of values of $\eta_\gamma^{[pair]}$ corresponding to a given bin
was estimated by producing $10^5$ sets of simulated data,
each obtained by reshuffling randomly the times of observation of the photons (of each GRB)
in our sample.
Of particular significance for our objective is the higher than expected observed
frequency of values of $\eta_\gamma^{[pair]}$ between 25 and 35.
Interestingly we find, using our simulated data obtained by time reshuffling,
that the excess in bin $25 \leq \eta_\gamma^{[pair]} \leq 35$ visible in Fig.2
is expected to occur accidentally only in 1.2$\%$ of cases.

\begin{figure}
\includegraphics[scale=0.6]{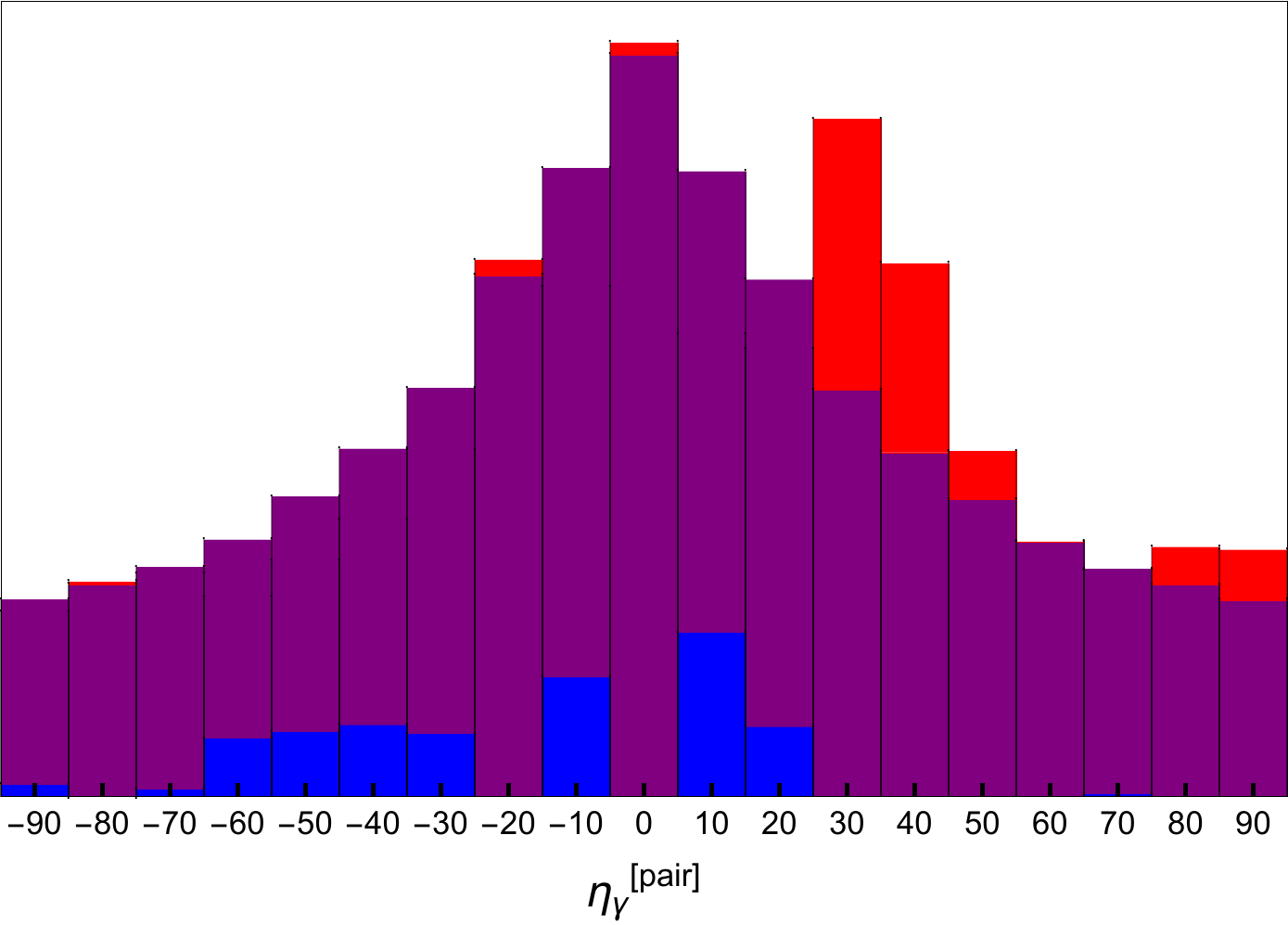}
\caption{Results of a study of the type already described in the previous Fig.2, but now
taking into account only pairs of photons that do not involve a ``high" photon. Color coding of the bars
is the same as for Fig.2.}
\end{figure}

In  Fig.5 we report the results of an analysis that is just like the analysis that produced Fig.2
but now excludes the contributions from the ``high" photons (with energy at emission greater than 40 GeV).
It is noteworthy that one still has a higher than expected observed
frequency of values of $\eta_\gamma^{[pair]}$ between 25 and 35, and for this case we estimate,
using our simulated data obtained by time reshuffling,
that the excess of occupancy of the bin
 $25 \leq \eta_\gamma^{[pair]} \leq 35$ visible in Fig.5
 should occur accidentally only in 0.6$\%$ of cases.

It is noteworthy that
a higher than expected observed
frequency of values of $\eta_\gamma^{[pair]}$ between 25 and 35 is present
also if we constrain the two photons in a pair to be of different type, for what concerns our
categories of ``high", ``medium"  and  ``low".
In Fig.6 we show the results we obtain for pairs composed of
a ``medium" ($15 GeV \leq E_0 \leq 40GeV$)
and a ``low" ($5 GeV \leq E_0 \leq 15GeV$) photon.
For this case we estimate,
using our simulated data obtained by time reshuffling,
that the excess of occupancy of the bin
 $25 \leq \eta_\gamma^{[pair]} \leq 35$ visible in Fig.6
 should occur accidentally only in 0.2$\%$ of cases.

\begin{figure}
\includegraphics[scale=0.6]{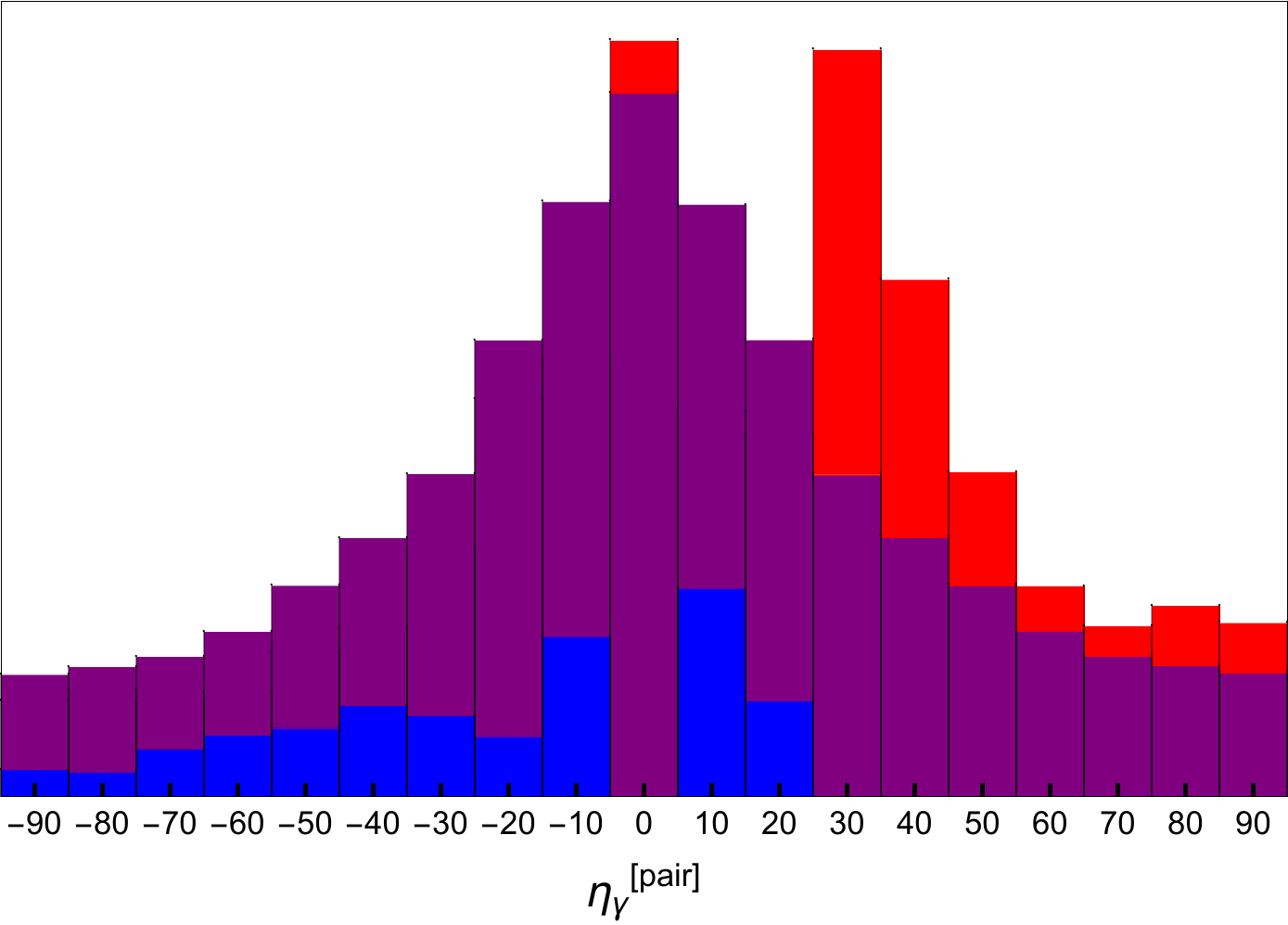}
\caption{Here we show the same type of results already shown in Figs.2 and 5, but now taking into
account only pairs composed of a ``medium"
and a ``low"  photon.}
\end{figure}

In Fig.7 we show the results we obtain for pairs composed of
a ``high" ($E_0 > 40GeV$)
and a ``low" ($5 GeV \leq E_0 \leq 15GeV$) photon.
As visible in Fig.7, once again we find
a higher than expected observed
frequency of values of $\eta_\gamma^{[pair]}$ between 25 and 35,
even though in this case the statistical significance is less striking:
using our simulated data obtained by time reshuffling,
we find that the excess of occupancy of the bin
 $25 \leq \eta_\gamma^{[pair]} \leq 35$ visible in Fig.7
 should occur accidentally in about 14$\%$ of cases
 (though this result reflects in part also the fact that we do not have
 high statistics of high-low pairs).

\begin{figure}
\includegraphics[scale=0.6]{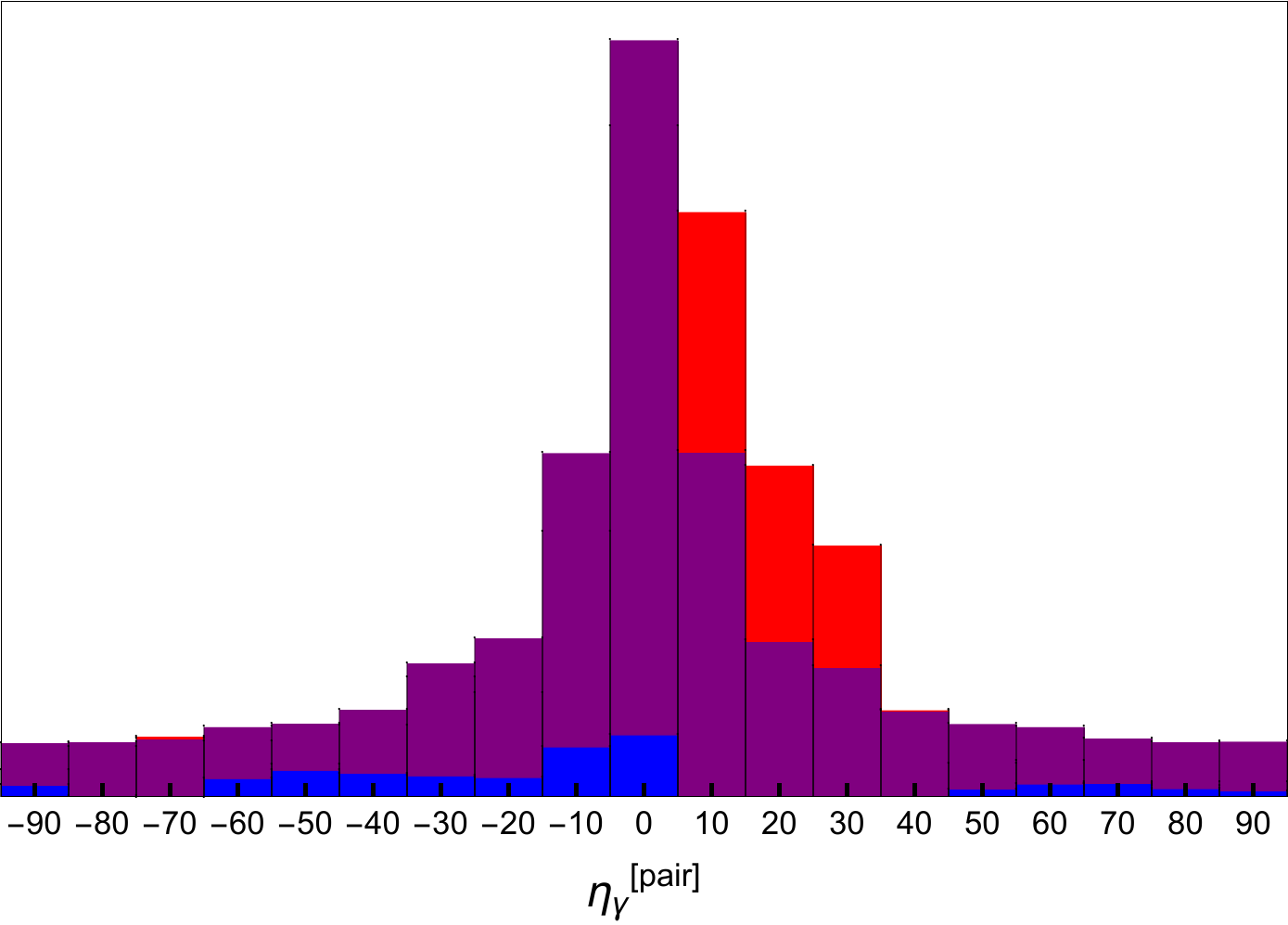}
\caption{Results of a study of the type already described in the previous Fig.2,5,6, but now we require the pair to be made of a ``high" and a ``low" photon.}
\end{figure}

In closing this appendix we go back to the content of our Fig.4, concerning triplets of photons.
We considered all triplets of photons (of course from the same GRB) in  our data set
and we assigned to each of these triplets a value of $\eta_\gamma$ obtained by performing the best linear fit with entries the observation times and the $E^*$ of the 3 photons, using equation
(\ref{maintwo}) (so the slope of the best-fit line going through the three points is $D(1) \eta_\gamma$).
For this triplet analysis the role of ``spurious results" (see the main text) can be stronger, and we tame
it by taking into account only values of $\eta_\gamma$ obtained by our best-linear-fit procedure
with $\chi^2$ smaller than 5.
The uncertainties in the energies are taken into account as done for the analyses based on pairs,
so here too a given triplet can contribute to more than one bin in our histogram.
Using our simulated data obtained by time reshuffling, we estimate
that the excess of occupancy of the bin
 $25 \leq \eta_\gamma^{[pair]} \leq 35$ visible in Fig.4
 should occur accidentally only in 0.3$\%$ of cases.

Between the main text and this appendix we discussed a total of 5 analyses which are to a large
extent independent, though not totally independent.
Each analysis uses different pairs, and in  one case
triplets, but for example the results reported in Fig.6 and Fig.7
could be used to anticipate to some extent the results of Fig.2 and Fig.4.
Considering the (rather high) level of independence of the different analyses it is striking
that in  all  cases we found an excess of results with $\eta_\gamma$
between 25 and 35. We  found that 4 of our analyses have significance between 0.2$\%$ and  1.2$\%$,
while the fifth analysis has significance of about 14$\%$.
The present data situation is surely intriguing, but dwelling on percentages is in our opinion premature.
We therefore prudently quote  in the main text an overall significance of about 0.5$\%$, but surely more refined
techniques of analysis of the overall statistical significance would produce an even more striking
estimate.

\end{document}